\documentclass[preprint]{aastex62-edited}

\usepackage{graphicx, natbib, verbatim, booktabs, threeparttable}
\usepackage{savesym}
\usepackage{amsmath}
\savesymbol{iint}         % restoring the amsmath version of the command
\usepackage{txfonts}
\restoresymbol{TXF}{iint} % and making the txfonts version of the command available as \TXFiint.

\begin{document}

% Unit vectors to be used
\newcommand{\omhat}{\hat{\Omega}}
\newcommand{\phat}{\hat{p}}
\newcommand{\nhat}{\hat{n}}
\newcommand{\mhat}{\hat{m}}
\newcommand{\uhat}{\hat{u}}
\newcommand{\vhat}{\hat{v}}
\newcommand{\xhat}{\hat{x}}
\newcommand{\yhat}{\hat{y}}
\newcommand{\zhat}{\hat{z}}

% Antenna response functions
\newcommand{\fplus}{F^{+}(\omhat)}
\newcommand{\fcross}{F^{\times}(\omhat)}

% GW metric tensors and amplitudes
\newcommand{\eplus}{e^+_{ab}(\omhat)}
\newcommand{\ecross}{e^\times_{ab}(\omhat)}
\newcommand{\hplus}{h_+(t,\omhat)}
\newcommand{\hcross}{h_\times(t,\omhat)}
\newcommand{\rplus}{r_+(t)}
\newcommand{\rcross}{r_\times(t)}

% Amplitudes of the pulsar timing residuals
\newcommand{\resamp}{\frac{\mathcal{M}^{5/3}}{D{\omega(t)}^{1/3}}} % evolving
\newcommand{\resampo}{\frac{\mathcal{M}^{5/3}}{D{\omega_0}^{1/3}}} % non-evolving

% Frequentist Earth-term statistic
\newcommand{\fe}{\mathcal{F}_e}

% Nicknames for frequent references
\defcitealias{libstempo}{\textsc{libstempo}}
\defcitealias{BabakEPTA}{B16}
\defcitealias{Ellis12}{E12}

\title{The Earth term in pulsar timing residuals is out of phase among the pulsars}

% https://journals.aas.org/aastex-v6-2-author-guide/#title_page
\author{Hyo Sun Park}
%\author[0000-0002-4986-0912]{Hyo Sun Park}
\affiliation{Department of Physics, Bryn Mawr College, 101 N Merion Ave, Bryn Mawr, PA, 19010, USA}
\affiliation{Department of Physics and Astronomy, Haverford College, 370 Lancaster Ave, Haverford, PA, 19041, USA}
\email{cpark2@brynmawr.edu}
%\author[0000-0003-4137-7536]{Andrea Lommen}
\author{Andrea Lommen}
\affiliation{Department of Physics and Astronomy, Haverford College, 370 Lancaster Ave, Haverford, PA, 19041, USA} 
\email{alommen@haverford.edu}

%\received{2021 June 7}
\date{2021 June 18}
\submitjournal{a NANOGrav Memorandum}

\shorttitle{The Earth term is out of phase among the pulsars}
\shortauthors{H. Park \& A. Lommen}
 
\begin{abstract}
    % Original version as of 4/21/2021.
    %We aim to resolve a misunderstanding about whether the so-called ``Earth term" in the pulsar timing response to a gravitational wave is in phase among a set of pulsars. We note that the misunderstanding has potentially arisen from the statements that the Earth term is ``coherent" or ``builds up coherently" among the pulsars. Using the pulsar timing residuals induced by a continuous gravitational wave, we show that the Earth term does not align across different pulsars except for the special case when the gravitational-wave source is edge-on, i.e. when the orbital inclination angle of the source is either $\iota=\pi/2$ or $3\pi/2$. We demonstrate the same concept using the pulsar timing software \citetalias{libstempo} by plotting the Earth terms from a set of pulsars.
    %We also point out that what authors mean by ``coherent" does not indicate that the Earth terms are in phase among the pulsars, but that they all share the same initial phase, $\Phi_0$.
    %{\color{blue} We also observe that the word ``coherent" can be interpreted in a few different ways. We clarify what authors mean by ``coherent" in those statements and point out that Earth terms being coherent does not indicate that the Earth terms are in phase among the pulsars.}
    
    % TODO: edited on 4/22/2021.
    We aim to resolve a misunderstanding about whether the so-called ``Earth term" in the pulsar timing response to a gravitational wave is in phase among a set of pulsars. We note that the misunderstanding has potentially arisen from the statements that the Earth term is ``coherent" or ``builds up coherently" among the pulsars. We clarify what authors mean by ``coherent" in these statements, pointing out that ``coherent" does not indicate that the Earth terms are in phase among the pulsars. Using the pulsar timing residuals induced by a continuous gravitational wave, we show that the Earth term does not align across different pulsars except for the special case when the gravitational-wave source is edge-on, i.e. when the orbital inclination angle of the source is either $\iota=\pi/2$ or $3\pi/2$. We demonstrate the same concept using the pulsar timing software \citetalias{libstempo} by plotting the Earth terms from a set of pulsars.
\end{abstract}


\section{Introduction}\label{sec:intro}
    
    %By monitoring a set of pulsars, the Pulsar Timing Arrays (PTAs) can detect gravitational-wave (GW) sources in the nano-Hz band \citep{M14}.
    The integral that accounts for the influence of a gravitational wave (GW) on the electromagnetic (EM) signal of the pulsar yields one piece at each of its endpoints: one that corresponds to the GW disturbance at the pulsar when the EM wave was emitted (the ``pulsar term"), and one that corresponds to the GW disturbance at the Earth when the EM wave was received (the ``Earth term") \citep{lommen15}. These two pieces in the residual response will in general be different in phase and frequency, and most GW detection strategies have to confront these differences. Many authors have made statements that for a given GW signal the Earth term is ``coherent" or ``builds up coherently" among a set of pulsars \citep[to name only a few]{EJM12, Ellis12, Perrodin18, Zhu16, BabakEPTA, lommen15, P12, SesanaVecchio2010}.
    The use of the word ``coherent," however, is ambiguous because these statements can be falsely interpreted as the Earth term being in phase across all the pulsars. Our goal in writing this paper is to resolve the misunderstanding these statements can cause.
    
    % Definition of "coherent radiation":
    % (Oxford) https://www.oxfordreference.com/view/10.1093/oi/authority.20110803095622393
    % (Wikipedia) In physics, two wave sources are perfectly coherent if their frequency and waveform are identical and their phase difference is constant. Coherence is an ideal property of waves that enables stationary interference.
    % (Britanica) a fixed relationship between the phase of waves in a beam of radiation of a single frequency
    There are several different meanings and uses of the word ``coherent."  One is a description of the radiation itself, another is a description of the analysis.
    We speak of two or more waves being coherent if the waves have the same frequency and constant phase differences between them \citep{OxfordDict}. % TODO: Check reference format
    For example, $\sin{(\omega t)}$ and $\cos{(\omega t)}$ are coherent as they have the same frequency, $\omega$, and a constant phase difference of $\pi/2$.
    Observe that coherent waves do not have to be in phase. However, waves with the same phase are often informally described as coherent, adding ambiguity in our use of the word. %Throughout this paper, when we describe a radiation being coherent, we will refer to the formal definition, where coherent waves can have a constant phase difference.
    
    ``Coherent" can also be used to describe the way we form a statistic.
    We speak of an analysis or a sum being coherent when we add the waves (or any data) together before squaring the sum of the waves \citep{Ellis12}. Observe that this allows the waves to interfere with each other both constructively and destructively. This is in contrast to an ``incoherent" sum in which we square each of the waves before adding them \citep{Ellis12}. Since each wave is squared, interference among the waves is not possible in incoherent analyses. We attempt to be very specific when we use the word ``coherent" so as to avoid any ambiguity.
    
    %We speak of two (or more) sets of waves being coherent if the waves have a constant phase relationship.  ``Coherent" does not technically mean that the waves have to have the same phase, although we tend to use it that way. This informal use is the first ambiguity in our use of the word, i.e. we tend to use it to describe two waves that have the same phase even though that does not have to be strictly true. We speak of analysis being coherent when we add the waves together before squaring them, vs an incoherent sum in which we square the waves before adding them. To summarize, ``coherent" can either describe the radiation itself, or the analysis.  If it describes the radiation it can mean either that the waves all have the same phase (informal use), or that they have a constant phase relationship (formal use.)  We attempt to be very specific when we use the word ``coherent" so as to avoid these ambiguities.
    
    % TODO: edited 4/22/2021
    Many of the aforementioned authors utilize in their GW detection strategies the fact that the Earth term is ``coherent" or ``builds up coherently" among the pulsars. In this paper, we point out that ``coherent" here does not imply that all the Earth terms have the same phase, and show that the Earth terms are generally out of phase among the pulsars. The Earth term is a combination of two terms: one is the contribution from the $+$-polarization, and the other from the $\times$-polarization of the gravitational wave.
    The $+$-polarization term is in phase across all the pulsars, and the $\times$-polarization term is in phase across all the pulsars. However, the combination of them is not, because the two terms have different relative amplitudes for different pulsars. This makes the Earth term have a different phase for each pulsar.\footnote{The situation is complicated by the fact that there are multiple definitions of the ``Earth term" in the literature. Our statement that the Earth term is generally out of phase is true if we define the Earth term as in this paper, as in Equation~\eqref{eqn:residual}. However, this statement does not hold for some definition of the Earth term. We discuss this in detail in Appendix~\ref{sec:appendix}.}
    %due to the antenna response functions by which they are multiplied prior to combining them.
    
    The paper is organized as follows. In Section~\ref{sec:review}, we review the mathematical expressions for the pulsar timing residuals induced by a continuous gravitational wave. In Section~\ref{sec:proof}, we show that the Earth terms line up only in the two special cases when the orbital inclination angle of the source is either $\iota=\pi/2$ or $3\pi/2$. We demonstrate the same results by plotting the Earth terms of $19$ different pulsars. Finally, we explain what authors mean by ``coherent" and conclude in Section~\ref{sec:conclusion}.

\section{Pulsar timing residuals induced by a continuous gravitational wave} \label{sec:review}

    %When a GW passes through space, it perturbs the spacetime between the pulsars and the Earth, and the times of arrival (TOAs) of the pulsar signals at the Earth deviate from the predicted model. The pulsar timing residual is the difference between the observed TOAs and the predicted TOAs. What we call the ``Earth term" and the ``pulsar term" arise as limits of integration of the frequency change over the path of the electromagnetic wave.  In this section we show this integral for a SMBHB and how the ``Earth term" and the ``pulsar term" arise in the pulsar timing response to the source. 

    We review the effect of a continuous gravitational wave emitted by a super-massive black hole binary (SMBHB) on the pulsar timing residuals. We assume that the SMBHB is non-spinning and in a circular orbit. We directly adopt the notations and calculations presented in \cite{Ellis12} (hereafter \citetalias{Ellis12}) and references therein. The gravitational wave perturbs the flat spacetime as it propagates, and the metric perturbation is expressed as:
    \begin{equation}
        h_{ab}(t) = \hplus\eplus + \hcross\ecross, \label{eqn:htt}
    \end{equation}
    where ``$ab$" represents the indices of the tensor, $\omhat$ the unit propagation vector pointing from the GW source to the solar system barycenter (SSB), and $h_{+,\times}$ the amplitudes of the $+$- and $\times$-polarizations of the GW, respectively. The polarization tensors $e^{+,\times}_{ab}$ are given by
    \begin{align}
        \eplus &= \mhat_a\mhat_b - \nhat_a\nhat_b, \\
        \ecross &= \mhat_a\nhat_b + \nhat_b\mhat_a.
    \end{align}
    The unit vectors introduced are defined in $\{\xhat,\,\yhat,\,\zhat\}$ by
    \begin{align}
        \omhat &= \{-\sin{\theta} \cos{\phi},\,
                    -\sin{\theta}\sin{\phi},\,
                    -\cos{\theta} \},             \\
        \mhat &= \{-\sin{\phi},\, \cos{\phi},\,0\},      \\
        \nhat &= \{-\cos{\theta}\cos{\phi}, \,
                    -\cos{\theta}\sin{\phi}, \,
                    \sin{\theta}\},
    \end{align}
    where $\theta,\,\phi$ are respectively the polar and azimuthal angles of the GW source location. The coordinate system $\{\xhat,\,\yhat,\,\zhat\}$ is defined such that the North celestial pole is in the $\zhat$-direction, the vernal equinox is in the $\xhat$-direction, and $\yhat = \zhat\times\xhat$. Note that $\mhat\times\nhat=-\omhat$ and that $\mhat$ points to increasing right ascension $\alpha$, and $\nhat$ points to increasing declination $\delta$.\footnote{$\{\alpha,\,\delta$\} and $\{\phi,\,\theta\}$ are related by $\phi=\alpha$ and $\theta=\pi/2 -\delta$.}
    Notice that we have two right-handed coordinate systems: $\{\xhat,\,\yhat,\,\zhat\}$ centered at the SSB and $\{\mhat,\,\nhat,\,-\omhat\}$ at the location of the GW source.
    
    To leading order and ignoring noise, the timing residual of a pulsar is written as:
    \begin{subequations}\label{eqn:residual}
    \begin{align}
        r(t,\omhat) &= r^p(t,\omhat)-r^e(t,\omhat) \\
                    &= \big[\fplus r_+(t_p)+\fcross r_\times(t_p)\big] -\big[\fplus r_+(t) + \fcross r_\times(t)\big],
    \end{align}
    \end{subequations}
    where $r^p$ (the first square-bracket) is the ``pulsar term" and $r^e$ (the second square-bracket) is the ``Earth term." Note that the Earth technically represents the SSB, as mentioned in \citetalias{Ellis12}. The antenna response functions $F^{+,\times}$, also known as the antenna pattern functions, of the given pulsar are defined as
    \begin{subequations}\label{eqn:antfns}
    \begin{align}
        \fplus &=  \frac{1}{2} \frac{(\phat\cdot\mhat)^2 -  (\phat\cdot \nhat)^2}{1 + \phat\cdot\hat{\Omega}}, \\
        \fcross &= \frac{(\phat\cdot\mhat) (\phat\cdot\nhat)}{1 + \phat\cdot\hat{\Omega}},
    \end{align}
    \end{subequations}
    % TODO--maybe: add sentences explaining or providing analogy of these antenna response functions --> Look at the textbook on Slack (file ending with *.ps) that Andrea sent me!
    where the unit vector $\phat$ points towards the pulsar from the Earth.
    
    % TODO: change explanation and the footnote
    In Equation~\eqref{eqn:residual}, $t$ denotes the time of observation at the Earth, and $t_p$, which is called the pulsar time, is given by
    \begin{equation}
        t_p = t - L(1 + \omhat\cdot\phat) \equiv t-\tau. \label{eqn:tp}
    \end{equation}
    The pulsar time $t_p$ always precedes the time $t$ at the Earth by $\tau$, where $\tau=L(1 + \omhat\cdot\phat)$ is the delay between two events: (1) the gravitational wave reaching the Earth, and (2) the information that the gravitational wave has reached the pulsar reaching the Earth \citep{lommen15}. Thus the pulsar term $r^p$ corresponds to an earlier epoch in the evolution of the GW source.\footnote{To give an analogy, imagine you get your news by two sources: one by social media and one by a print newspaper delivered to your doorstep. The print newspaper is going to carry news from a time $\tau$ earlier, where $\tau$ represents the lag between information getting posted on social media and information arriving on your doorstep via the newspaper. If you look at social media and the print newspaper at the same time, your brain will be carrying news from two different times: news(now) + news(now - $\tau$).}
    However, notice that $t_p$ does not exactly equal the time when the radio pulse was emitted from the pulsar. In fact, $t_p$ equals the time of radio emission from the pulsar, i.e. $t_p = t-L$, only when $\omhat$ is perpendicular to $\phat$.
    If $\omhat=\phat$, for example, i.e. when the Earth is in between the pulsar and the GW source, then Equation~\eqref{eqn:tp} gives $t_p=t-2L$.
    
    \begin{comment}
    {\color{red}
    where $\tau=L(1 + \omhat\cdot\phat)$ is the delay between the two events: (1) the gravitational wave reaching the Earth, and (2) the information that the gravitational wave has reached the pulsar reaching the Earth \citep{lommen15}.
    Note that the pulsar time $t_p$ always precedes the time $t$ at the Earth by $\tau$, and thus the pulsar term $r^p$ corresponds to an earlier epoch in the evolution of the GW source.\footnote{To give an analogy, if you get your news by two sources, one by social media and one by the print newspaper delivered to your doorstep, the print newspaper is going to carry news from a time $\tau$ earlier, where $\tau$ represents the lag between information getting posted on social media and information arriving on your doorstep via the newspaper. If you look at social media and the print newspaper at the same time, your brain will be carrying news from two different times: news(now) + news(now - $\tau$).}
    The pulsar term is frequently described as the gravitational wave arriving at the pulsar, but this description is misleading because $t_p$ does not exactly equal the time when the radio pulse was emitted from the pulsar. In fact, $t_p$ equals the time of radio emission from the pulsar, i.e. $t_p = t-L$, only when $\omhat$ is perpendicular to $\phat$.
    However, if $\omhat=\phat$, i.e. when the Earth is in between the pulsar and the GW source, then Equation~\eqref{eqn:tp} gives $t_p=t-2L$.
    }
    \end{comment}

    Finally, $r_{+,\times}$ in Equation~\eqref{eqn:residual} are given by
    \begin{align}
        r_+ (t) &= \resamp \big[-(1+\cos^2\iota)\cos{2\psi}\sin2(\Phi(t)-\phi_n) -
                    2\cos\iota\sin2\psi\cos2(\Phi(t)-\phi_n)\big],   \label{eqn:rplus} \\
        r_\times(t) &= \resamp\big[-(1+\cos^2\iota)\sin{2\psi}\sin2(\Phi(t)-\phi_n) +
                    2\cos\iota\cos2\psi\cos2(\Phi(t)-\phi_n)\big].   \label{eqn:rcross}
    \end{align}
    Here, $\iota$ is the inclination angle of the SMBHB orbital plane, $\psi$ the GW polarization angle, and $D$ the luminosity distance to the SMBHB.
    The chirp mass $\mathcal{M}$ of the binary is defined as $\mathcal{M} = (m_1 m_2)^{3/5}/(m_1+m_2)^{1/5}$, where $m_1,\,m_2$ represent the masses of the two black holes in the SMBHB.
    $\phi_n$ is the orbital phase at the line of nodes, which is defined as the intersection of the SMBHB orbital plane with the tangent plane of the sky \citep{Wahl87}.
    The orbital phase and frequency of the SMBHB are given respectively by
    \begin{align}
        \Phi(t) &= \Phi_{0} + \frac{1}{32\mathcal{M}^{5/3}}\left({\omega_{0}}^{-5/3} - \omega(t)^{-5/3}\right), \label{eqn:orbphase}\\
        \omega(t) &= \omega_{0}\left(1-\frac{256}{5}\mathcal{M}^{5/3}{\omega_{0}}^{8/3}t\right)^{-3/8},\label{eqn:orbfreq}
    \end{align}
    where $\Phi_0$ and $\omega_0$ are respectively the initial orbital phase and frequency determined at the time when the observation begins.\footnote{$\Phi_0$ and $\omega_0$ are related to the initial GW phase and frequency by $\Phi_0=\Phi_{gw,0}/2$ and $\omega_0 = \omega_{gw,0}/2$.}
    
    It is important to note that different authors define their Earth term differently. As we will see in Section~\ref{sec:proof}, the Earth term, $r^e$, does not generally align among the pulsars. However, there is another definition of the Earth term in the literature that does align across all the pulsars. We compare different definitions of the Earth term in the literature in Appendix~\ref{sec:appendix}. In this paper, we follow the notations used in \citetalias{Ellis12} and thus define $r^e$, given in Equation~\eqref{eqn:residual}, as the Earth term.

\section{The Earth terms are Not in phase}\label{sec:proof}

    In this section, we show that the Earth term is generally out of phase among a set of pulsars. The Earth terms align only when the orbital inclination angle of the SMBHB is either $\iota=\pi/2$ or $3\pi/2$, i.e. when the source is edge-on. We demonstrate the same results by plotting the Earth terms of $19$ different pulsars using the pulsar timing software \citetalias{libstempo} \citep{libstempo}.\footnote{The \citetalias{libstempo} library can be found at: \url{https://github.com/vallis/libstempo}.} The Earth term $r^e$ can be fully written as (Equations~\ref{eqn:residual}, \ref{eqn:rplus}, and \ref{eqn:rcross}):
    \begin{align}
        r^e(t,\omhat) = -\resamp \bigg\{&\fplus \big[-(1+\cos^2\iota)\cos{2\psi}\sin 2\Phi'(t)
                        - 2\cos\iota\sin2\psi\cos 2\Phi'(t)\big] \nonumber\\
                        + &\fcross \big[-(1+\cos^2\iota)\sin{2\psi}\sin 2\Phi'(t)
                        + 2\cos\iota\cos2\psi\cos 2\Phi'(t)\big] \bigg\}, \label{eqn:proof1}
    \end{align}
    where $\Phi'(t) = \Phi(t)-\phi_n$. We simplify this equation employing the following trick. Since $\sin 2\Phi'(t)$ and $\cos 2\Phi'(t)$ are sinusoids of the same frequency $\omega(t)$ (Equation~\ref{eqn:orbfreq}), the two terms inside the first square-bracket can be combined into a single sinusoid with some new amplitude $A(\iota,\psi)$ and new phase $\Phi_A(t)$. Notice that the frequency $\omega(t)$ remains the same, but the new phase $\Phi_A(t)$ depends on the ratio of the two coefficients $-(1+\cos^2\iota)\cos{2\psi}$ and $-2\cos\iota\sin2\psi$. Using the same method, we combine the two terms inside the second square-bracket, producing a new amplitude $B(\iota,\psi)$ and a new phase $\Phi_B(t)$. This new phase $\Phi_B(t)$ is dependent on the ratio of the coefficients $-(1+\cos^2\iota)\sin{2\psi}$ and $2\cos\iota\cos2\psi$. Then, Equation~\eqref{eqn:proof1} is reduced to:
    \begin{align}
        r^e(t,\omhat) = -\resamp \bigg\{\fplus A(\iota,\psi)\sin\Phi_A(t) + \fcross B(\iota,\psi)\cos\Phi_B(t)\bigg\}. \label{eqn:proof2}
    \end{align}
    
    Here, we see that the Earth term is a sum of two sinusoidal waves, $\sin\Phi_A(t)$ and $\cos\Phi_B(t)$, of the same frequency, $\omega(t)$. Therefore, the two terms in Equation~\eqref{eqn:proof2} can \textit{again} be combined into a single sinusoid with the same frequency, but with a new amplitude and a new phase that depends on the two coefficients $\fplus A(\iota,\psi)$ and $\fcross B(\iota,\psi)$.
    However, note that $A(\iota,\psi)$ and $B(\iota,\psi)$ are the same for all the pulsars, and so are $\Phi_A(t)$ and $\Phi_B(t)$. The only things that differ between the pulsars are the antenna response functions $F^{+,\times}$, which depend on the sky location, $\phat$, of each pulsar (see Equation~\ref{eqn:antfns}).
    %Since the new phase is dependent on the ratio of $\fplus A(\iota,\psi)$ and $\fcross B(\iota,\psi)$,
    This means that the overall phase of this single sinusoid, $r^e(t,\omhat)$, will be determined by the relative amplitudes of $\fplus$ and $\fcross$. Hence, it follows that the Earth terms will have different phases for different pulsars.
    
    There are two special cases when the Earth terms line up amongst the pulsars. This is when the second terms in both square-brackets in Equation~\eqref{eqn:proof1} go to zero, i.e. when $\iota =\pi/2$ or $3\pi/2$ (when the SMBHB is edge-on). In these cases, we have $\cos\iota=0$, and Equation~\eqref{eqn:proof1} becomes:
    \begin{align}
        r^e(t,\omhat)\bigg|_{\iota=\pi/2,\,3\pi/2} &= \resamp\bigg\{\fplus\cos{2\psi}\sin 2\Phi'(t) + \fcross\sin{2\psi}\sin 2\Phi'(t)\bigg\} \nonumber \\
                    &= \resamp\bigg(\fplus\cos{2\psi} + \fcross\sin{2\psi}\bigg)\sin 2\Phi'(t).\label{eqn:proof3}
    \end{align}
    Thus, when $\iota =\pi/2$ or $3\pi/2$, the Earth terms from all the different pulsars will have the same phase, $2\Phi'(t)$, though their amplitudes will be different because the factor $\fplus\cos{2\psi} + \fcross\sin{2\psi}$ varies among the pulsars due to $F^{+,\times}$.
    
    One might wonder if, for an arbitrary $\iota$, we can choose a $\psi$ such that all the Earth terms line up. However, there is no such value of $\psi$. For all the Earth terms to align, we need a condition where the phase of the Earth term does not depend on the ratio of $F^{+}$ and $F^{\times}$, as in Equation~\eqref{eqn:proof3}.
    This condition is achieved only when the $\sin2\Phi'(t)$ terms in both square-brackets go to zero, or when the $\cos2\Phi'(t)$ terms in both square-brackets vanish in Equation~\eqref{eqn:proof1}.
    To eliminate any of the terms, we need to choose $\psi=0,\,\pi/2,\, \pi,\,3\pi/2,$ or $2\pi$. However, observe that none of these makes the Earth terms align amongst the pulsars. 
    To see this clearly, consider the following example. When $\psi=0$, the $\cos2\Phi'(t)$ term vanishes from the first square-bracket and the $\sin2\Phi'(t)$ term vanishes from the second square-bracket. So Equation~\eqref{eqn:proof1} becomes:
    \begin{align}
        r^e(t,\omhat)\bigg|_{\psi=0} =\resamp \bigg\{\fplus(1+\cos^2\iota)\cos{2\psi}\sin 2\Phi'(t) - \fcross(2\cos\iota)\cos2\psi\cos 2\Phi'(t)\bigg\}. \label{eqn:proof4}
    \end{align}
    Notice that we have basically obtained the same situation as in Equation~\eqref{eqn:proof2}. If we combine the two terms in Equation~\eqref{eqn:proof4}, the phase of the resultant sinusoid will depend on the ratio of $\fplus$ and $\fcross$, making the Earth terms have a different phase for each pulsar. We acquire similar results for $\psi=\pi/2,\,\pi,\, 3\pi/2$, and $2\pi$. 
    Hence, we conclude that there is no value of $\psi$ that makes the Earth terms line up across all the pulsars.
    %But in these cases, $\psi$ can only make either the $\sin2\Phi'(t)$ term in the first square-bracket and the $\cos2\Phi'(t)$ term go to zero, or the $\cos2\Phi'(t)$ term in the first square-bracket and the $\sin2\Phi'(t)$ term go to zero.
    %This is because $\cos2\psi$ is multiplied by $\sin2\Phi'(t)$ in the first square-bracket, while $\sin2\psi$ is multiplied by $\sin2\Phi'(t)$ in the second square-bracket (and vice versa for $\cos2\Phi'(t)$).
    
\begin{figure}
        \centering
        \includegraphics[scale=0.35]{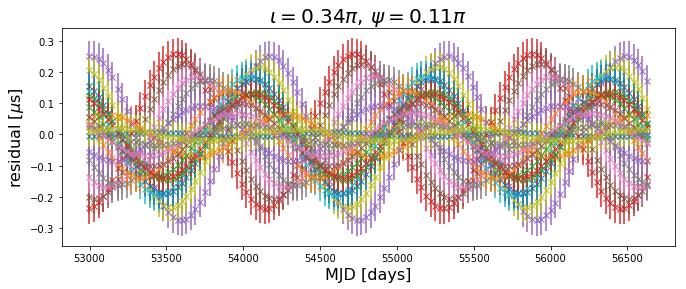}
        \includegraphics[scale=0.35]{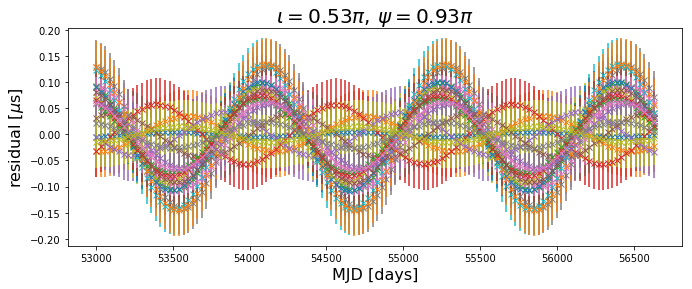}
        \includegraphics[scale=0.35]{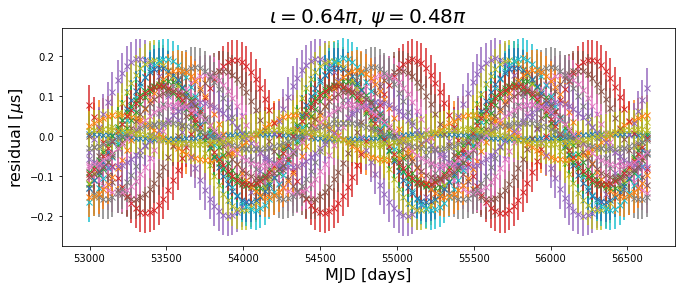}
        \includegraphics[scale=0.35]{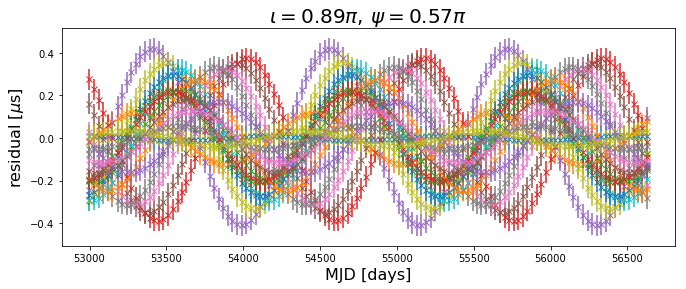}
        \caption{\footnotesize The Earth terms from $19$ different pulsars with four different random inclination ($\iota$) and polarization ($\psi$) angles. Each color represent a single pulsar. The Earth terms are clearly out of phase among the pulsars. However, the Earth terms do align when $\iota=\pi/2$ or $3\pi/2$, i.e. when the source is edge-on, as shown in Figure~\ref{fig:earth-incpi2}.}
        \label{fig:earthrandom}
    \end{figure}

\begin{figure}
        \centering
        \includegraphics[scale=0.35]{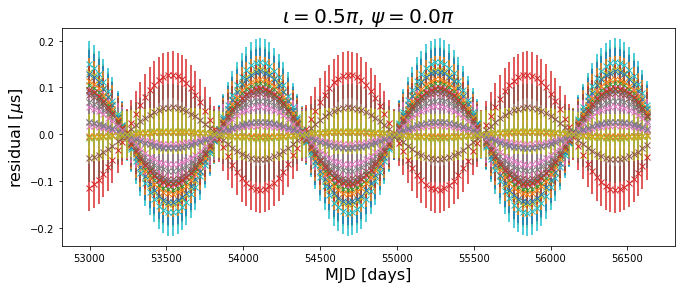}
        \includegraphics[scale=0.35]{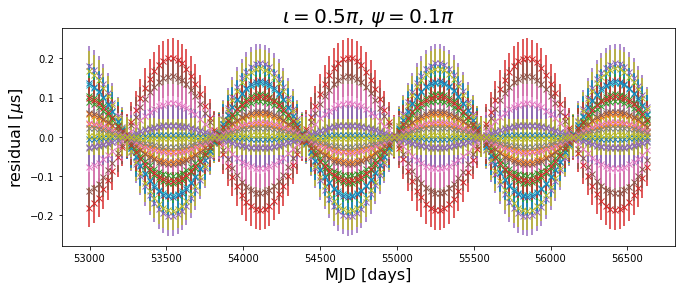}
        \includegraphics[scale=0.35]{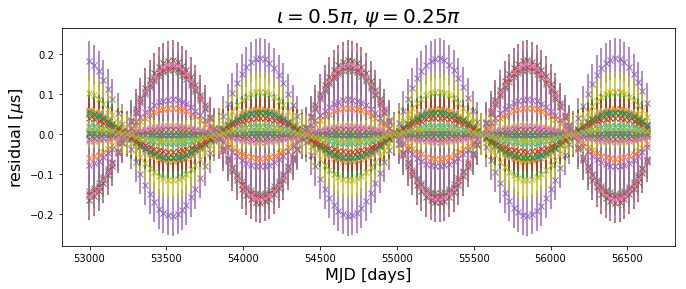}
        \includegraphics[scale=0.35]{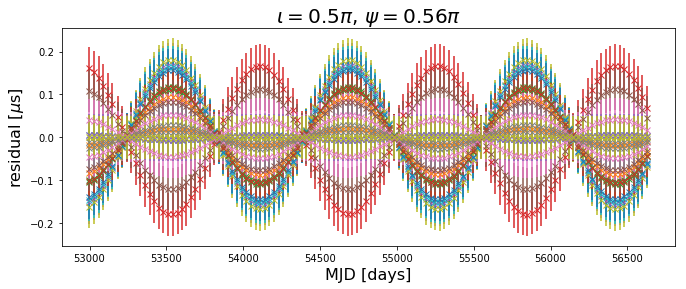}
        \includegraphics[scale=0.35]{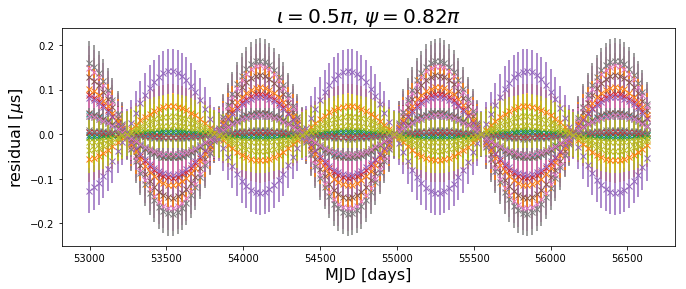}
        \includegraphics[scale=0.35]{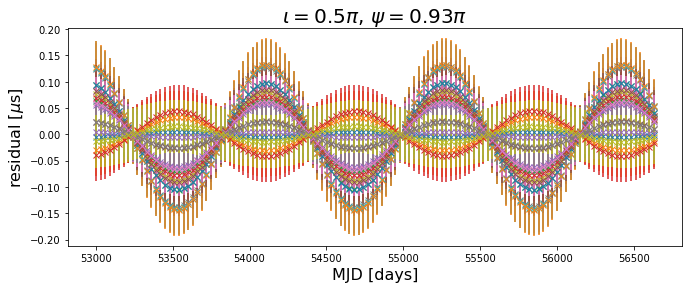}
        \caption{\footnotesize The Earth terms from $19$ different pulsars with the inclination angle of $\pi/2$ (edge-on) and six different random polarization angles. Each color represent a single pulsar. This is one of the two special cases ($\iota =\pi/2$ or $3\pi/2$) where the Earth terms line up across all the pulsars. General cases when $\iota \not=\pi/2,3\pi/2$ are shown in Figure~\ref{fig:earthrandom}. Note that the Earth terms for some pulsars have opposite signs because $F^{+,\times}$ can produce an overall negative sign depending on the pulsar location. The Earth terms have different amplitudes for different pulsars because $F^{+,\times}$ vary among the pulsars (see Equation~\ref{eqn:proof3}).}
        \label{fig:earth-incpi2}
    \end{figure}

    We demonstrate our results using \citetalias{libstempo}'s function \texttt{add\_cgw} with a set of $19$ different pulsars.\footnote{The \citetalias{libstempo} library can be found at: \url{https://github.com/vallis/libstempo}.} The function \texttt{add\_cgw} in the \texttt{toasim} package injects a continuous GW emitted by a SMBHB into the timing residuals. It offers the option to exclude the pulsar term (by setting the parameter \texttt{psrTerm=False}), allowing us to extract only the Earth term from the timing residuals.
    In Figure~\ref{fig:earthrandom}, the Earth terms from $19$ different pulsars are plotted for random values of $\iota$ and $\psi$. We clearly see that the Earth terms are out of phase among the pulsars.
    On the other hand, when the inclination angle is either $\iota =\pi/2$ or $3\pi/2$ (when the system is edge-on), all the Earth terms line up, as shown in Figure~\ref{fig:earth-incpi2}.
    Notice that the Earth terms have different amplitudes for different pulsars due to the factor $\fplus\cos{2\psi} + \fcross\sin{2\psi}$ in Equation~\eqref{eqn:proof3}.
    Also, the Earth terms for some pulsars have an overall negative sign from other Earth terms. This is because $F^{+,\times}$ can be either positive or negative depending upon each pulsar's direction $\phat$ (see Equations~\ref{eqn:antfns} and~\ref{eqn:proof3}).
    % some pulsars still have an overall negative sign in their residuals
    %Note that the Earth terms from some pulsars have opposite signs because  F+ and Fx can be positive or negative depending upon the pulsar direction. 
    
\section{Conclusion} \label{sec:conclusion}
    % TODO: whole section edited 4/22/2021.
    We pointed out that our ambiguous use of the word ``coherent" could lead to the misconception that the Earth terms in the pulsar timing residuals are in phase across all the pulsars.
    As we showed in Section~\ref{sec:proof}, however, the Earth terms are generally out of phase. % could lead one to falsely presume that ... However, the word ``coherent" can be interpreted in several different ways.
    
    The difference in phase among the Earth terms arises from the antenna response functions, $F^{+, \times}$, which depend on the position, $\phat$, of a particular pulsar.
    The Earth term is a combination of two sinusoidal terms: one corresponding to the +-polarization which is multiplied by $\fplus$ and the other corresponding to the $\times$-polarization which is multiplied by $\fcross$. These two terms have the same frequency, so we can combine them into a single Earth-term sinusoid which has the same frequency but a new amplitude and a new phase. The phase of this resultant sinusoid is dependent on the relative amplitudes of the two original sinusoidal terms. In particular, the new phase depends on the ratio of $\fplus$ and $\fcross$, but $F^{+, \times}(\omhat)$ vary among the pulsars. Therefore, the Earth term has a different phase for each pulsar. The Earth terms align only when the inclination of the SMBHB orbital plane is either $\iota=\pi/2$ or $3\pi/2$, i.e. when the binary system is edge-on. We demonstrated our results by plotting the Earth terms from $19$ different pulsars, using the pulsar timing software \citetalias{libstempo} (see Figures~\ref{fig:earthrandom} and~\ref{fig:earth-incpi2}).
    
    A few GW detection techniques, for example the frequentist Earth-term $\fe$-statistic, exploit the fact that the Earth term in the pulsar timing response to a gravitational wave is ``coherent" or ``builds up coherently" among a set of pulsars (e.g. \citetalias{Ellis12}; \citealt{BabakEPTA}, hereafter \citetalias{BabakEPTA}). We stress that the word ``coherent" in these statements does not imply that the Earth term is in phase among the pulsars.
    %Rather, the authors \citetalias{Ellis12} and \citetalias{BabakEPTA} mean two things: first, there is only one phase parameter, $\Phi_0$, that needs to be maximized over for all the different pulsars; second, in doing the analysis the Earth terms are summed before they are squared (as opposed to squared before they are summed).
    In fact, Earth terms being ``coherent" and Earth terms ``building up coherently" mean two different things. The former is a description of the radiation, i.e. the Earth-term wave itself, whereas the latter is a description of the analysis used in the $\fe$-statistic (see Section~\ref{sec:intro} for different meanings of the word ``coherent").
    
    In the maximum likelihood calculation for $\fe$ in \citetalias{Ellis12} and \citetalias{BabakEPTA}, the initial phase, $\Phi_0$, of the binary system (Equation~\ref{eqn:orbphase}) is one of the four parameters, $(\mathcal{M}^{5/3}D^{-1}, \,\iota, \,\Phi_0,\, \psi)$, over which the authors analytically maximize their statistic in order to search for a GW signal. By Earth terms being ``coherent," \citetalias{Ellis12} and \citetalias{BabakEPTA} mean that the parameter that refers to the initial phase, $\Phi_0$, of the binary is the same in all the Earth terms such that one needs to maximize over only one phase parameter for all the different pulsars. On the other hand, when the authors say that the Earth terms ``build up coherently" among the pulsars, they are describing the way the Earth-term signals are combined in the calculation of the $\fe$-statistic. The authors are making the point that the $\fe$-statistic is a coherent statistic because in doing the analysis the Earth terms are summed before they are squared (as opposed to squared before they are summed).
    We refer the reader to~\citetalias{Ellis12} for complete mathematical treatment of the $\fe$-statistic.

\begin{acknowledgements}
    We thank Jeffrey S. Hazboun, Maura McLaughlin, Chiara M. F. Mingarelli, Joseph D. Romano, and Caitlin A. Witt for valuable comments and discussions. This work was supported by the NANOGrav NSF Physics Frontier Center.
\end{acknowledgements}

\begin{appendix}
    \section{Different Definitions of the Earth term}\label{sec:appendix}
        In Section~\ref{sec:proof}, we showed that the Earth term in the pulsar timing residuals is generally out of phase across the pulsars. However, there are multiple definitions of the ``Earth term" in the literature, and whether the Earth term is in phase depends upon one's definition of the Earth term. In this Appendix, we compare two different definitions of the Earth term used in the literature and show that the statement that the Earth terms are generally out of phase does not hold true for the latter definition.
        
        Some authors incorporate the antenna response functions $F^{+,\times}$ (Equation~\ref{eqn:antfns}) in their Earth term so that the Earth term represents one component of the timing residual that is actually measured at the Earth (ignoring noise), as defined in this paper. Many authors including \citetalias{Ellis12}, \citetalias{BabakEPTA}, \cite{cc10}, \cite{Perrodin18}, and \cite{SesanaVecchio2010} have adopted this definition. This is the Earth term $r^e$ we use, given in Equation~\eqref{eqn:residual}:
        \begin{align}
            r^e(t,\omhat) &= -\fplus r_+(t) - \fcross r_\times(t). \label{eqn:earthtermresp}
        \end{align}
        
        On the other hand, some define $r_{+,\times}(t)$ to be their Earth term (Equations~\ref{eqn:rplus} and~\ref{eqn:rcross}). Note that $\rplus$ and $\rcross$ are the ingredients in the calculation of $r^e$ as shown in Equation~\eqref{eqn:earthtermresp}, but they do not yield a measurable quantity until they are multiplied by $F^{+,\times}$. A number of authors including \cite{Jenet04} and \cite{Arz20} have used this latter definition. Note that unlike $r^e$ which is generally out of phase, each of $\rplus$ and $\rcross$ is in phase across all the pulsars because $r_{+,\times}(t)$ depend solely on the properties of the GW source (e.g. $\iota,\,\psi,\,\Phi,\,\phi_n$).
        Hence, the statement that the Earth term is generally out of phase amongst the pulsars does not hold true for this latter definition.
        
        %We also note that there are some authors who define the GW strain at the Earth to be their ``Earth term" \citep[e.g.][]{Anh09, EJM12}.
\end{appendix}

\bibliography{ms}

\end{document}